\documentclass[12pt]{article}
\usepackage[T1]{fontenc}
\usepackage{bm}
\usepackage{cite}
\usepackage{mathrsfs}
\usepackage{slashed}
\usepackage{graphicx} 
\usepackage{hyperref}
\usepackage{verbatim}
\usepackage{color}
\usepackage{enumerate}
\usepackage{authblk}
\usepackage[hang,flushmargin]{footmisc}
\usepackage{lipsum}
\usepackage[font=footnotesize]{caption}
\usepackage[utf8]{} 
\usepackage{empheq}
\usepackage{amssymb,amsmath,amsfonts}
\usepackage[makeroom]{cancel}
\usepackage{hyperref}
\usepackage{color}

﻿
﻿
\numberwithin{equation}{section}

\usepackage{physics}

﻿
\definecolor{blue-violet}{rgb}{0.54, 0.17, 0.89}
\definecolor{PineGreen}{cmyk}{0.92, 0, 0.59, 0.25}
\definecolor{OliveGreen}{cmyk}{0.64, 0, 0.95, 0.40}
\definecolor{RawSienna}{cmyk}{0, 0.72, 1, 0.45}
\definecolor{Gray}{cmyk}{0, 0, 0, 0.50}
\definecolor{MidnightBlue}{cmyk}{0.98, 0.13, 0, 0.43}
\definecolor{Orange}{cmyk}{0, 0.61, 0.87, 0}
\definecolor{LimeGreen}{cmyk}{0.50, 0, 1, 0}
\definecolor{Green}{cmyk}{1, 0, 1, 0}

﻿
﻿

﻿
﻿
﻿

﻿
﻿
\renewcommand{\tilde}{\widetilde}
\renewcommand{\hat}{\widehat}

\usepackage{mathrsfs}

﻿
﻿
﻿
\begin{document}
﻿
\title{\!\!\LARGE{Generalized Schwarzian thermodynamics \\ from gravitational perturbations of near-extremal Kerr}\vspace{10pt}}

﻿
\author[a]{\normalsize{Marcela C\'ardenas}\footnote{\href{mailto:marcela.cardenasl@uss.cl}{marcela.cardenasl@uss.cl}}}
\author[b]{\normalsize{Benjam\'in Cuevas-Tapia}\footnote{\href{mailto:benjamin.cuevas@uc.cl}{benjamin.cuevas@uc.cl}}}
\author[c]{\normalsize{Jessica Trespalacios}\footnote{\href{mailto:jessica.trespalacios@udec.cl}{jessica.trespalacios@udec.cl}}}
\affil[a]{\footnotesize\textit{Universidad San Sebasti\'an, Facultad de Ingenier\'ia, Bellavista 7, Recoleta, Santiago, Chile.}}
\affil[b]{\footnotesize\textit{Facultad de F\'isica, Pontificia Universidad Cat\'olica de Chile, Vicu\~na Mackenna 4860, Macul, Santiago, Chile.}}
\affil[c]{\footnotesize\textit{Departamento de Ingenier\'ia Matem\'atica, Facultad de Ciencias F\'isicas y Matem\'aticas, Universidad de Concepci\'on, Concepci\'on, Chile.}}
\date{}
﻿
\maketitle
﻿
\begin{abstract}
We derive the generalized Schwarzian thermodynamics of near-extremal Kerr directly from four-dimensional gravitational perturbations of the near-horizon geometry. We construct two families of exact linear perturbations of the near-NHEK background: a non-normalizable mode that deforms the asymptotic boundary, and a normalizable mode supported near the horizon. Both perturbations are smooth solutions of the vacuum Einstein equations reduced to the Jackiw--Teitelboim equation for the dual scalar, without conical singularities. We study their asymptotic symmetries and show that they generate an algebra of time reparametrizations; requiring the gauge parameters to depend on the Schwarzian mode, the isometry condition selects the stationary KdV hierarchy. The same hierarchy is independently enforced by the vacuum Einstein equations on the perturbation coefficient. In the Euclidean regime, we evaluate the Hamiltonian action and show that each mode independently generates a tower of odd-power corrections to the near-extremal entropy.
\end{abstract}
﻿
\newpage
﻿
﻿
\section{Introduction} 

The near-horizon region of extremal black holes is the natural arena where
the infrared dynamics of gravity becomes tractable. For the extremal Kerr
black hole, the decoupling limit yields the Near-Horizon Extreme Kerr (NHEK)
geometry \cite{Bardeen:1999px}, a fibration of a U(1) circle over AdS$_2$
whose isometry group is enhanced to SL$(2,\mathbb{R})\times\mathrm{U}(1)$.
This enhancement underlies the Kerr/CFT proposal \cite{Guica:2008mu} and
its extensions to near-extremal configurations \cite{Bredberg:2009pv}.
However, the rigid AdS$_2$ throat does not support finite-energy excitations:
any perturbation that carries energy into the throat back-reacts and destroys
its asymptotic structure \cite{Dias:2009ex,Amsel:2009ev,Breitenlohner:1982jf}. Moving slightly
away from extremality therefore requires breaking the conformal
symmetry in a controlled way.

The universal mechanism for this breaking is captured by nearly-AdS$_2$
holography and Jackiw--Teitelboim (JT) gravity \cite{Jackiw:1984je,Teitelboim:1983ux}.
The reparametrizations of the boundary time are spontaneously and explicitly
broken down to SL$(2,\mathbb{R})$, and their low-energy dynamics is governed
by the Schwarzian action \cite{Almheiri:2014cka,Jensen:2016pah,Maldacena:2016upp,Engelsoy:2016xyb}. These edge modes of the throat control the
thermodynamics of near-extremal black holes at low temperatures, including
the quantum corrections that resolve the apparent breakdown of the
semiclassical description
\cite{Iliesiu:2020qvm,Kapec:2023ruw,Rakic:2023vhv,Banerjee:2023quasilocal,Arnaudo:2024,Kapec:2024law}. For Kerr, the near-horizon region was perturbed by a JT sector in
\cite{Castro:2019crn,Castro:2021csm}, where gravitational perturbations
of NHEK were organized in terms of the JT equation; recently,  more arguments have appear in favour that JT gravity is the effective theory that dominates
at low temperature \cite{Castro:2025itb}.  On the other hand,  in the perturbation analyses of
\cite{Castro:2021csm,Mariani:2025kgw}, the JT sector is sourced
by conical singularities of the four-dimensional geometry.

The connection between KdV integrability and black hole thermodynamics was
first uncovered in three-dimensional gravity. In \cite{Perez:2016vqo}, it
was shown that AdS$_3$ gravity with KdV boundary conditions admits an
infinite set of commuting conserved charges, and that the resulting
thermodynamics organizes into a generalized Gibbs ensemble with odd-power
corrections to the entropy. This structure was subsequently linked to a bigger 
integrable asymptotic symmetry algebra of AdS$_3$ \cite{Cardenas:2021vwo,Cardenas:2025qqi},
and its two-dimensional avatar---the generalized Schwarzian theory with KdV
symmetry-breaking---was developed in \cite{Cardenas:2024hah}, where it was
shown to govern the thermodynamics of nearly-AdS$_2$ black holes with an
infinite tower of KdV deformations. The present work establishes that this
integrable structure is not merely a feature of lower-dimensional models:
it is encoded directly in the four-dimensional near-horizon geometry of
near-extremal Kerr.

In this article we realize this mechanism explicitly: starting from the
integrable structure of the near-NHEK symmetry algebra, we construct two
families of exact gravitational perturbations and derive their thermodynamic
contributions. Our starting point is
the observation that the infinitesimal symmetries of any locally AdS$_2$
geometry fibred as in NHEK are generated by a scalar dual to the Killing
vector, which satisfies precisely the JT equation. Rather than sourcing this
scalar with a conical singularity, we promote it to a genuine dynamical field
and construct two families of exact linear perturbations of the near-NHEK
geometry that are smooth solutions of the vacuum Einstein equations: a
\emph{non-normalizable} mode $\mathfrak{b}_+=\mathcal{B}_+(t)/r^2$ that
deforms the asymptotic boundary, and a \emph{normalizable} mode
$\mathfrak{b}_-=\mathcal{B}_-(t)\,r^2$ localized near the horizon. These
two profiles are related by the inversion $r\mapsto 1/r$ of the AdS$_2$
isometry group, and they carry complementary information: $\mathcal{B}_+$
is the source seen from the asymptotic boundary, while $\mathcal{B}_-$ is
the response seen from the horizon.

When the symmetry parameters are allowed to depend locally on the
Schwarzian mode $\mathcal{L}(t)$, the Killing equations reproduce the
stationary Korteweg--de Vries (KdV) hierarchy
\cite{Gervais:1985fd,Perez:2016vqo,Cardenas:2024hah}. In this language the
near-NHEK black hole is a particular soliton of the KdV equation, and the
near-horizon region is endowed with an infinite set of conserved integrals
of motion. The two modes carry different conformal weights under boundary
reparametrizations: $\mathcal{B}_+$ transforms as a weight-one object and
$\mathcal{B}_-$ as a weight-two density. 

In the Euclidean regime we evaluate the Hamiltonian action within the
Regge--Teitelboim formalism \cite{Regge:1974zd,Gibbons:1976ue}. The asymptotic Killing
vectors associated with $\mathfrak{b}_+$ generate a surface term at the
outer boundary proportional to the KdV Hamiltonians, reproducing the
generalized Schwarzian thermodynamics of \cite{Cardenas:2024hah} with the
coupling fixed by the angular momentum; the resulting corrections to the
near-extremal entropy are odd powers of the temperature. The near-horizon Killing vectors associated with $\mathfrak{b}_-$ generate a
finite charge at the horizon with the same structure as the asymptotic charge
of $\mathfrak{b}_+$.  Because $\mathfrak{b}_+$ and $\mathfrak{b}_-$ correspond to alternative perturbations,  with their own asymptotic behaviour,  they define
independent theories, each with their own generalized Gibbs ensemble, yet both analysis yields an independent tower of odd-power corrections,
\begin{equation*}
\log Z^{(\pm)} \simeq S_0 - 8\pi\beta\,c_\phi\,J + \sum_j \nu_j^\pm\,C_j^\pm\,T^{2j-1},
\end{equation*}
where $\nu_j^+$ are the KdV chemical potentials sourced by $\mathfrak{b}_+$ at
$r\to\infty$, and $\nu_j^-$ those sourced by $\mathfrak{b}_-$ at $r=r_h$. 

The article is organized as follows. Section~\ref{Gauge transformations via Scalar-Duality and KdV symmetries} reviews the
symmetries of the near-horizon geometry and their reformulation in terms of
the dual scalar.
Section~\ref{Near-NHEK and KdV perturbations} constructs the two families
of gravitational perturbations.
Section~\ref{Hamiltonian} computes the asymptotic and near-horizon Killing
vectors, evaluates the Hamiltonian action in the Euclidean regime, and
derives the thermodynamics at the saddle point. We close with our
conclusions.
\section{Symmetries of the near-horizon geometry: Gauge transformations via scalar duality} \label{Gauge transformations via Scalar-Duality and KdV symmetries}

The NHEK geometry \cite{Bardeen:1999px} is the vacuum solution
\begin{equation}\label{NHEK}
    ds^2= \Lambda(\theta)\left[-R^2 dT^2+\frac{dR^2}{R^2}+d\theta^2\right]+\Gamma(\theta)\left[d\Phi+RdT\right]^2 ,
\end{equation}
with angular functions
\begin{equation}\label{angular-functions}
\Lambda(\theta)=M^{2}(1+\cos^{2}\theta)\qquad\text{and}\qquad \Gamma(\theta)=\frac{4M^{2}\sin^{2}\theta}{1+\cos^{2}\theta},
\end{equation}
where $M$ is the mass of the extremal Kerr black hole from which it is obtained by a decoupling limit. The $(T,R)$ sector is an AdS$_{2}$ space, the coordinate $\Phi$ parametrizes a U(1) fibre with connection $R\,dT$, and surfaces of constant $\theta$ are warped AdS$_{3}$ geometries \cite{Anninos:2008fx,Compere:2008cv}. The isometry group is SL$(2,\mathbb{R})\times U(1)$, generated by
\begin{equation}\label{Killin-Sl2R}
\zeta = \qty(\epsilon(T) + \frac{\epsilon''(T)}{2R^2})\partial_T - R\epsilon'(T)\partial_R
- \frac{\epsilon''(T)}{R}\partial_\Phi + c_\Phi\partial_\Phi,\qquad \epsilon'''(T)=0 ,
\end{equation}
so that $\epsilon(T)=c_{-1}+c_{0}T+c_{1}T^{2}$ and $c_\Phi$ is a constant. The vectors \eqref{Killin-Sl2R} close the Lie algebra
\begin{equation}\label{Lie-bracket}
\{\zeta[\epsilon_1],\zeta[\epsilon_2]\}^{\mu} = \zeta^{\mu}[\epsilon_3]
\qquad \text{for}\qquad  \epsilon_{3}=\epsilon_1\epsilon_2'-\epsilon_2\epsilon_1',
\end{equation}
and the generators $L_{n}\equiv\zeta[T^{n+1}]$,
\begin{equation}\label{SL(2,R)-generators}
L_{-1}=\partial_{T}, \qquad  L_{0}=T\partial_{T}-R\partial_{R}, \qquad L_{1}=\left(T^2+\frac{1}{R^2}\right)\partial_{T}-2TR\,\partial_{R} -\frac{2}{R}\,\partial_{\Phi} ,
\end{equation}
satisfy $\{L_{n},L_{m}\}=(m-n)L_{n+m}$ for $n,m=-1,0,1$, and commute with the U(1) generator $\partial_\Phi$.
The global SL$(2,\mathbb{R})$ can be enhanced to the full group of time reparametrizations by allowing the AdS$_{2}$ throat to fluctuate, as in nearly-AdS$_{2}$ geometries, which are responsible for driving black holes away from extremality. Following \cite{Kapec:2019hro}, we take the improper coordinate transformation\footnote{This coordinate change is inspired by the three-dimensional construction of \cite{Roberts:2012aq}.}
\begin{equation}\label{nonlinear coord-change}
    T=f+\frac{2 f'^{2} f''}{4 r^{2} f'^{2}-f''^{2}},\qquad
    R=\frac{4 r^{2} f'^{2}-f''^{2}}{4 r f'^{3}},\qquad
    \Phi=\phi+\log\left[\frac{2rf'-f''}{2rf'+f''}\right],
\end{equation}
where $f=f(t)$ and primes denote derivatives with respect to the new time, $\partial_{t}=(\,)'$. It maps \eqref{NHEK} into
\begin{equation}\label{4dmetric}
ds^2 = \Lambda(\theta)\qty(-\qty(r + \frac{\mathcal{L}(t)}{2r})^2\dd t^2 + \frac{\dd r^2}{r^2}+ \dd\theta^2) + \Gamma(\theta)\qty(\dd\phi + \qty(r - \frac{\mathcal{L}(t)}{2r})\dd t)^2,
\end{equation}
where
\begin{equation}\label{Schwarzian}
 \mathcal{L}(t)= \left\{f,t\right\}=\frac{f'''}{f'}- \frac{3}{2}\left(\frac{f''}{f'}\right)^{2}
\end{equation}
is the Schwarzian derivative, invariant under $f\to (af+b)/(cf+d)$ with $ad-bc=1$. The family \eqref{4dmetric} contains NHEK for $f(t)=t$, i.e. $\mathcal{L}=0$, and the near-NHEK black hole \cite{Bredberg:2009pv} for $f(t)=e^{\kappa t}$, i.e. $\mathcal{L}=-\kappa^{2}/2$, with horizon at $r=\kappa/2$. The latter describes a closer zoom into the horizon of a slightly non-extremal Kerr black hole, where $\kappa$ measures the deviation from extremality, with Hawking temperature $T_{H}=\lambda\kappa/(4\pi M^{2})+\mathcal{O}(\lambda^{2})$ in terms of the decoupling parameter $\lambda\to0$ \cite{Bredberg:2009pv}.

Following \cite{Castro:2021csm}, we consider the geometry
\begin{equation}\label{4dmetric-general}
ds^2 = \Lambda(\theta)\qty(\gamma_{ab}\dd x^a \dd x^b + \dd\theta^2) + \Gamma(\theta)\qty(\dd\phi + \mathcal{A}_a\dd x^a)^2,
\end{equation}
 where $x^a=(t,r)$,  $\Lambda,\Gamma$ are the angular functions defined in \eqref{angular-functions}, $\gamma_{ab}$
is now \textit{any} locally AdS$_{2}$ metric and $\mathcal{A}=\mathcal{A}(x^a)$ is a vector field. 
%It is is possible to show vacuum Einstein equations with vanishing cosmological constant,  and consequently,  Ricci flat solutions,
%\begin{equation}
%R_{\mu\nu}=0
%\end{equation}
%are solved by 2-dimensional spacetimes $\gamma$ and the additional condition for the vector field $\mathcal{A}_{a}$ satisfying
With this ansatz, the vacuum Einstein equations with vanishing cosmological constant 
$$R_{\mu\nu} = 0$$ 
reduce to 
the conditions
\begin{equation}\label{2Dcurvature}
R(x^{a})=-2,\qquad \epsilon^{ab}\partial_{a}\mathcal{A}_{b}=-1.
\end{equation}
Here $R(x_a)$ is the curvature from the two-dimensional $\gamma_{ab}$ metric and $\epsilon^{ab}$ denotes the Levi-Civita tensor with sign convention $\epsilon_{tr}=\sqrt{-\det \gamma}$. 

We look for symmetries of this general geometry \eqref{4dmetric-general}.  We start by proposing the vector field 

$\zeta^{\mu}=\{\bar{\zeta}^{a}(x^{a},\phi),0,\zeta^{\phi}(x^{a},\phi)\},$
i.e., the vector field has no azimuthal component and is independent of $\phi$. 
%where it has been considered no azimuthal dependence on its components,  nor on its direction. 
From the equation $\delta g_{\phi\phi}=\mathcal{L}_{\zeta}g_{\phi\phi}=0$, one gets the condition 
\begin{equation}\label{condition1}
\partial_{\phi}\zeta^{\phi}+\mathcal{A}_{a}\partial_{\phi}\bar{\zeta}^{a}=0.
\end{equation}  

The equation
$\delta g_{a\phi}=\mathcal{L}_{\zeta}g_{a\phi} $,
reads
\begin{equation}\label{Lie phi phi}
\delta g_{a\phi}=\Gamma (\theta) \left( \mathcal{L}_{\bar{\zeta}}\mathcal{A}_{a}+\mathcal{A}_{a}(\partial_{\phi}\zeta^{\phi}+ \mathcal{A}_{b}\partial_{\phi}\bar{\zeta}^{b} )+ \partial_{a}\zeta^{\phi}\right)+\Lambda(\theta)\gamma_{ab}\partial_{\phi}\bar{\zeta}^b. 
\end{equation}
Now we insert \eqref{condition1},  reducing \eqref{Lie phi phi} to %the above to
\begin{equation}
\Gamma (\theta)\delta \mathcal{A}_{a}=\Gamma (\theta)(\mathcal{L}_{\bar{\zeta}}\mathcal{A}_{a}+ \partial_{a}\zeta^{\phi})+\Lambda(\theta)\gamma_{ab}\partial_{\phi}\bar{\zeta}^b.
\end{equation}

Since only $\Gamma(\theta)$ and $\Lambda(\theta)$ depend on $\theta$ and these are linearly independent functions, this gives the pair of equations
\begin{equation}\label{variation A}
\delta \mathcal{A}_{a}=\mathcal{L}_{\bar{\zeta}}\mathcal{A}_{a}+ \partial_{a}\zeta^{\phi},\qquad \gamma_{ab}\partial_{\phi}\bar{\zeta}^b=0.
\end{equation}
The remaining variation is given by $\delta g_{ab}=\mathcal{L}_{\zeta} g_{ab}=\Lambda\delta \gamma_{ab}+\Gamma \delta (\mathcal{A}_{a}\mathcal{A}_{b})$, that decouples into the variation of the 2-dimensional metric and the gauge field $\mathcal{A}$. Considering again the linear independence of $\Lambda$ and $\Gamma$ and \eqref{variation A} we get, 
 \begin{equation}
 \delta \gamma_{ab}=\nabla_{a}\bar{\zeta}_{b}+\nabla_{b}\bar{\zeta}_{a}.
 \end{equation}
Finally,  the equation $\mathcal{L}_{\zeta}g_{\theta\theta}=0,$ is identically satisfied.
As a summary, we have the set of constraint equations
 \begin{eqnarray}
\partial_{\phi}\zeta^{\phi}+\mathcal{A}_{a}\partial_{\phi}\bar{\zeta}^{a}&=&0,\\
\gamma_{ab}\partial_{\phi}\bar{\zeta}^b=\partial_{\phi}\bar{\zeta}_{a}&=&0,
 \end{eqnarray}
and the set of gauge transformations
 \begin{eqnarray}
 \delta \gamma_{ab}&=&\nabla_{(a}\bar{\zeta}_{b)},  \\
\delta \mathcal{A}_{a}&=&\mathcal{L}_{\bar{\zeta}}\mathcal{A}_{a}+ \partial_{a}\zeta^{\phi}.
 \end{eqnarray}
 From the above set of transformations, it becomes clear that part of the symmetry transformations along $\mathcal{A}_{a}$ can be recovered from the ones associated to the 2-dimensional metric. Moreover,  they can be rewritten considering that $\bar{\zeta}^{a}=\epsilon^{ab}\partial_{b}\Phi$ and contracting the transformation with the antisymmetric tensor,
 \begin{eqnarray}
  \epsilon^{ac}  \delta \gamma_{ab}&=&\gamma^{cd}(2\nabla_{b}\nabla_{d}\Phi-\gamma_{bd}\Box \Phi), \\\label{delta g}
  \delta \mathcal{A}_{a}&=&-\partial_{a}(\Phi+\mathcal{A}_{b}\epsilon^{bc}\partial_c\Phi+\zeta^{\phi}). \label{delta A}
 \end{eqnarray}
To obtain the equation \eqref{delta A}, we have considered the equation of motion $\partial_{a}\mathcal{A}_{b}-\partial_{b}\mathcal{A}_{a}=-\epsilon_{ab}$, which is simply a recast of the second equation in \eqref{2Dcurvature}.
Here we focus on the fact that  the symmetries of the full fibration \eqref{4dmetric-general}
can be encoded in a single scalar function. Following \cite{Castro:2021csm},
one can show that the Killing equations
$\mathcal{L}_\zeta g_{\mu\nu}=0$ for the ansatz \eqref{4dmetric-general},
with $\zeta^\mu$ independent of $\phi$, reduce after separation of the
angular dependence to the two-dimensional conditions
\begin{equation}\label{delta A'}
\nabla_a\nabla_b\Phi = \tfrac{1}{2}\gamma_{ab}\,\Box\Phi,\qquad
\zeta^\phi = -\Phi - \mathcal{A}_b\epsilon^{bc}\partial_c\Phi + \Phi_0,
\end{equation}
where $\Phi=\Phi(x^a)$ is a scalar, $\Phi_0$ is an integration constant, and
the two-dimensional component of the Killing vector is recovered as
$\bar\zeta^a=\epsilon^{ab}\partial_b\Phi$. The first equation in \eqref{delta A'}
is precisely the Jackiw--Teitelboim equation: the Hessian of $\Phi$ is pure
trace on the AdS$_2$ background $\gamma_{ab}$, so $\Phi$ plays the role of
the JT dilaton.

For the near-NHEK metric \eqref{4dmetric}, the solution of \eqref{delta A'}
renders
\begin{equation}\label{Phi-solution}
\Phi = r\,\epsilon(t) + \frac{\mathcal{L}(t)\,\epsilon(t) - \epsilon''(t)}{2r},
\end{equation}
with $\epsilon(t)$ an arbitrary function of time. The $\zeta^\phi$ component
then yields the transformation law of the Schwarzian mode,
\begin{equation}\label{D1}
\delta_\epsilon\mathcal{L} = \mathcal{D}_1\epsilon,\qquad
\mathcal{D}_1 \equiv \partial_t^3 + 2\mathcal{L}\,\partial_t + \mathcal{L}',
\end{equation}
which is the action of a Virasoro coadjoint vector on a stress tensor of
central charge by restoring units through $\mathcal{L}\rightarrow\frac{k}{4\pi}\mathcal{L}$ with central charge $c=\frac{4\pi}{k}$. The transformation of $\Phi$ under a second Killing
parameter $\epsilon_2$,
\begin{equation}\label{translaw-gauge}
\delta_{\epsilon_2}\Phi[\epsilon_1] = \Phi[\epsilon_1'\epsilon_2 - \epsilon_1\epsilon_2'],
\end{equation}
reproduces the Lie bracket \eqref{Lie-bracket}. 

\section{Gravitational perturbations}

In this section we construct two families of exact linear perturbations of the (near)-NHEK background using the gauge formulation of constant-curvature spacetimes. The key feature is that the perturbation can be recast as the Jackiw--Teitelboim equation, so the KdV hierarchy emerges directly from the vacuum Einstein equations. The two radial profiles have complementary fall-offs: one decays faster than the background as $r\to\infty$, while the other is well-defined near $r\to 0$.

\subsection{Near-NHEK as a solution of the KdV equation and integrable perturbations} \label{Near-NHEK and KdV perturbations}
To propose new gravitational perturbations, we draw on the equations \eqref{2Dcurvature}. First, we construct a locally AdS$_{2}$ spacetime $\gamma_{ab}$ using the first-order formulation of the constant-curvature equation in 2D \cite{Fukuyama:1985gg,Isler:1989hq,Chamseddine:1989yz}, which has been extensively used in the literature to construct 2D asymptotic conditions \cite{Grumiller:2015vaa,Grumiller:2017qao,Godet:2020xpk}. Here, its dynamics can be recast in terms of the JT scalar. In a second step, this enables us to construct a 4D spacetime using the ansatz \eqref{4dmetric-general}, demanding the radial dependence to decay faster than the near-NHEK background.

\subsubsection{Locally AdS$_{2}$ metric}
We rewrite the constant curvature equation $R[\gamma]=-2$ in terms of the one-form connection $\mathbf{A}$,
\begin{equation} \label{eom-zero-curvature}
\mathcal{F} =d\mathbf{A}+\mathbf{A}\wedge \mathbf{A}=0,
\end{equation}
being $\mathcal{F}$ the field strength. The equation \eqref{eom-zero-curvature} encompass the first order equations  
\begin{equation}
de_{c}+\epsilon_{cd}\omega e^{d}=0\qquad\text{and}\qquad  d\omega+\frac{1}{2}\epsilon_{cd}e^{c}e^{d}=0,
\end{equation}
by means of the relationship with the first order fields $e,\omega$,
\begin{equation}\label{1st order fields}
\mathbf{A}_{I}=(e_{0},e_{1},\omega)  
\end{equation}
 for $\{I=c,2\}$\footnote{Here $c,d,\dots$ represent indices in the algebra, which should not be confused with the latin indices in the $(t,r)$ coordinates showed in subsection Section~\ref{Gauge transformations via Scalar-Duality and KdV symmetries}.}, where $\epsilon^{012}=1$ is the antisymmetric invariant tensor.  Here the relevant algebra is 
$\mathfrak{
so}(2,1)$,
\begin{equation}
 \left[T_{I},T_{J}\right]=\eta^{KL}\epsilon_{IJ}T_{L}   
\end{equation}
in which the fields are spanned.  Additionally,  the associated invariant bilinear form is $\eta_{IJ}=2\langle T_{I},T_{J}\rangle= \operatorname{diag}(1,1,-1)$ and the metric components in terms of the vielbein can be retrieved using the formula 
\begin{equation}\label{metric-vielbein} 
    \gamma_{\mu\nu}=\eta_{cd}e^{c}_{\mu}e^{d}_{\nu}=2 \rm \langle e_{\mu}e_{\nu}\rangle.
\end{equation}
It is convenient to work in the $\mathfrak{sl}(2,\mathbb{R})$ basis, as it is isomorphic to $\mathfrak{so}(2,1)$.  Being $L_{n}$ with $n=\left\{-1,0,1 \right\}$ the generators of $\mathfrak{sl}(2,\mathbb{R})$,  the transformations among the two basis are given by
\begin{equation}\label{Transformation-basis}
 T_{2}=\frac{1}{2}\left(L_{-1}+L_{1}\right),\text{\ensuremath{\quad}}T_{0}=\frac{1}{2}\left(L_{-1}-L_{1}\right),\quad T_{1}=L_{0}.   
\end{equation}
Their algebra satisfy the commutation relation
$\left[L_{n},L_{m}\right]=\left(n-m\right)L_{n+m}$, and the only non-zero components of the invariant bilinear form are given by $\langle L_{0}, L_{0}\rangle=1/2$ and $\langle L_{1},L_{-1}\rangle=-1$ \footnote{
The chosen matrix representation for the $\mathfrak{sl}(2,\mathbb{R})$ generators is,
\begin{equation*}
L_{-1}=\begin{pmatrix}0 & 0\\
1 & 0
\end{pmatrix}\quad,\quad L_{0}=\begin{pmatrix}-\frac{1}{2} & 0\\
0 & \frac{1}{2}
\end{pmatrix}\quad,\quad L_{1}=\begin{pmatrix}0 & -1\\
0 & 0
\end{pmatrix}\;.\label{PTT-sl(2,R)-MR}
\end{equation*}}.
 
Using \eqref{1st order fields} that relate the $\mathfrak{so}(2,1)$ components of vielbein and the spin connection with the gauge fields  $\mathbf{A}=\mathbf{A}^{+}L_{1}+\mathbf{A}^{-}L_{-1}+\mathbf{A}^{0}L_{0}$ and the transformation \eqref{Transformation-basis},  one can stablish the relationship 
\begin{eqnarray}\label{vielbeins}
    e^{0}=\mathbf{A}^{+}-\mathbf{A}^{-},\qquad e^{1}= \mathbf{A}^{0},\qquad \omega=-(\mathbf{A}^{+}+\mathbf{A}^{-}).
\end{eqnarray}
Here we propose the conditions
\begin{equation}\label{gauge-connections}
\mathbf{A}_{t}=-r L_{1}+\frac{\mathcal{L}}{2r}L_{-1}\qquad \mathbf{A}_{r}=\mathfrak{b}L_{1}+\mathfrak{c}L_{-1}-\mathfrak{a}L_0
\end{equation}
where $\mathfrak{a}=\mathfrak{a}(x^a)$, $\mathfrak{b}=\mathfrak{b}(x^{a})$ and $\mathfrak{c} = \mathfrak{c}(x^a)$ are free functions which are constrained by constant curvature equation \eqref{eom-zero-curvature}.  A similar gauge choice was previously considered in \cite{Cardenas:2024hah} to recover nearly AdS$_{2}$ spacetimes.  Indeed,  analizing the $L_{0}$ and $L_{-1}$ components of \eqref{eom-zero-curvature}, one get the equations, 
\begin{equation}\label{function-conditions}
\mathfrak{a}= -\frac{\mathfrak{b}'+ 1}{r},\qquad \mathfrak{c}= -\frac{\mathfrak{b}'' + \mathcal{L} \mathfrak{b}}{2r^2},
\end{equation}
that relate the functions $\mathfrak{a},\mathfrak{c}$ with $\mathfrak{b}$. Employing now the relationship between the 2D metric and the gauge field components \eqref{metric-vielbein}, leads to the following ansatz for any locally AdS$_{2}$ geometry \footnote{To recover a Lorentzian signature in the metric,  one should additionally make the simultaneous identifications $\mathbf{A}_{t}\rightarrow i\mathbf{A}_{t}$, $\mathfrak{b}\rightarrow i\mathfrak{b}$.},
\begin{align}\label{new-2D-metric}
 \gamma_{tt} &= -\qty(r + \frac{\mathcal{L}}{2r})^2,\notag \\
 \gamma_{tr} &= \qty(r + \frac{\mathcal{L}}{2r}) (\mathfrak{b} - \mathfrak{c}),\\\label{new-2D-metric}
 \gamma_{rr} &= \mathfrak{a} ^2 - (\mathfrak{b} - \mathfrak{c})^2,\notag
\end{align}
Using \eqref{new-2D-metric} and \eqref{function-conditions}, the constant negative curvature equation $R[\gamma]=-2$ is solved by,
\begin{equation}
2\mathcal{L} \mathfrak{b}'+\mathcal{L}' \mathfrak{b}+\mathfrak{b}'''=0.
\end{equation}
As one can observe, the equation is only characterized by temporal derivatives of $\mathfrak{b}$ whereas the radial dependence is left unconstrained. We will fix the radial dependence in the next subsection, depending on the features of the  fall-off perturbation.

\subsubsection{Radial redundancy and the two independent modes}
The freedom in the radial profile of $\mathfrak{b}$ can be seen as reflection of a deeper symmetry of AdS$_2$. The isometry group $\mathrm{SL}(2,\mathbb{R})$ acts on the Poincar\'{e} coordinate $z=1/r$ by inversions $r\mapsto 1/r$, which interchanges the two independent power-law solutions of the radial part of the gauge connection. Concretely, adopting the Poincar\'{e} coordinate $z=1/r$, the two linearly independent radial profiles
\begin{equation}\label{two-modes}
\mathfrak{b}_+ = \frac{\mathcal{B}_+(t)}{r^2}, \qquad \mathfrak{b}_- = \mathcal{B}_-(t)\,r^2
\end{equation}
are mapped into each other by the inversion $r\mapsto 1/r$, confirming that the radial coordinate is redundant up to this $\mathrm{SL}(2,\mathbb{R})$ freedom. In the standard AdS$_2$/CFT$_1$ language \cite{Klebanov:1999tb,Ishibashi:2004wx,Marolf:2006nd}, $\mathfrak{b}_+$ has conformal dimension $\Delta_+=2$ (non-normalizable mode, source) while $\mathfrak{b}_-$ has conformal dimension $\Delta_-=-2$ (normalizable mode, response or VEV), with the two related by the inversion symmetry.

Both profiles satisfy the same temporal equation \eqref{eom-KdV-4D} independently, so the full two-parameter family
\begin{equation}\label{full-b}
\mathfrak{b}(r,t) = \frac{\mathcal{B}_+(t)}{r^2} + \mathcal{B}_-(t)\,r^2
\end{equation}
is a solution for any pair $(\mathcal{B}_+,\mathcal{B}_-)$ satisfying the KdV hierarchy equation 
$$
2\mathcal{L} \mathcal{B}'+\mathcal{L}'\mathcal{B}+\mathcal{B}'''=0.$$
\subsubsection{Perturbed near-NHEK}

Using \eqref{2Dcurvature} and \eqref{4dmetric-general}, we aim to construct a spacetime of the form,
\begin{equation}\label{pertubed-NHEK}
\bar{g}_{\mu\nu}=g_{\mu\nu}+h_{\mu\nu},
\end{equation}
where $g_{\mu\nu}$ is the background spacetime associated to the near-NHEK solution \eqref{4dmetric} including the Schwarzian mode,  and $h_{\mu\nu}$ is an axisymmetric perturbation. 

We take the locally AdS$_{2}$ spacetime found in the previous subsection \eqref{new-2D-metric} and exploit the two independent radial profiles \eqref{two-modes} to construct two distinct classes of gravitational perturbation.

\subsubsection*{Asymptotic perturbation ($\mathfrak{b}_+$).}
We first select the non-normalizable profile $\mathfrak{b}=\mathcal{B}_+(t)/r^{2}$, which ensures that the deviation from the NHEK geometry decays faster than the background as $r\rightarrow \infty$. Both perturbations constructed below are smooth vacuum solutions with no conical singularities at the poles \cite{Mariani:2025kgw}. This choice determines the vector $\mathcal{A}$ via \eqref{2Dcurvature},
\begin{equation}\label{condition A}
\mathcal{A}_{t}=r+\frac{-\mathcal{L} \left(\mathcal{B}_+'+3 r^2\right)}{6 r^3},\qquad \mathcal{A}_{r}=-\frac{\mathcal{B}_+}{r^2},
\end{equation}
and leads to the perturbation (renaming $\mathcal{B}_+\to\mathcal{B}$ for brevity),
\begin{eqnarray}
h_{tt}&=&\Gamma(\theta)\left( -\frac{ \mathcal{B}' \mathcal{L}}{3r^{2}}+\frac{ \mathcal{B}' \mathcal{L}^{2}}{6r^4}+\frac{(\mathcal{B}')^{2} \mathcal{L}^2}{36 r^6}\right)\nonumber\\
h_{tr}&=& \Gamma(\theta)  \left(-\frac{\mathcal{B}}{r}+\frac{\mathcal{B} \mathcal{L}}{2 r^3}+\frac{\mathcal{B}'\mathcal{B} \mathcal{L}}{6 r^5}\right)+\Lambda(\theta) \left(\frac{\mathcal{B}}{r}+\frac{\frac{1}{2} \mathcal{B}''+\mathcal{B} \mathcal{L}}{r^3}+\frac{\mathcal{L} \left(\mathcal{B}''+\mathcal{B}\mathcal{L}\right)}{4 r^5}\right)\nonumber\\
 h_{t\theta}&=&0\qquad\qquad h_{t\varphi}=-\Gamma(\theta) \frac{ \mathcal{B}' \mathcal{L}}{6r^3} \label{perturbation}\\
h_{rr}&=&\Gamma(\theta)\frac{\mathcal{B}^2}{r^4}+\Lambda(\theta) \left(\frac{(\mathcal{B}')^2}{r^6}-\frac{\left(\mathcal{B}''+2 \mathcal{B} \left(\frac{1}{2} \mathcal{L}+r^2\right)\right)^2}{4 r^8}\right)\nonumber \\
h_{r\theta}&=&0\qquad\qquad h_{r\varphi}=-\Gamma(\theta) \frac{\mathcal{B}}{r^{2}}\qquad\qquad  h_{\theta\theta}=h_{\theta\varphi}=h_{\varphi\varphi}=0.\nonumber
\end{eqnarray}
All components decay as $r\to\infty$, and the perturbation vanishes when $\mathcal{B}=0$.

\subsubsection*{Near-near-horizon perturbation ($\mathfrak{b}_-$).}
The complementary radial mode $\mathfrak{b}=\mathcal{B}_-(t)\,r^{2}$ vanishes at $r=0$ and grows toward the asymptotic boundary. We refer to it as the \emph{near-near-horizon} perturbation because it is regular at the near-NHEK horizon, in analogy with the near-horizon character of the NHEK geometry itself relative to the full Kerr spacetime. In the AdS$_2$/CFT$_1$ dictionary, $\mathfrak{b}_-$ is the \emph{normalizable} mode in the sense that its coefficient $\mathcal{B}_-(t)$ is not freely prescribed as an asymptotic boundary condition but is instead determined dynamically as the response (VEV) of the dual operator. The two modes $\mathfrak{b}_\pm$ play complementary roles: $\mathfrak{b}_+$ sources the operator from the boundary at $r\to\infty$, while $\mathfrak{b}_-$ encodes the near-horizon response. Substituting $\mathfrak{b}=\mathcal{B}_-(t)\,r^2$ into \eqref{function-conditions} gives
\begin{equation}
\mathfrak{a} = -2\mathcal{B}_- - \frac{1}{r}, \qquad
\mathfrak{c} = -\frac{\mathcal{B}_-\mathcal{L}}{2} - \frac{\mathcal{B}_-}{r^2},
\end{equation}
and the vector $\mathcal{A}$, determined to linear order in $\mathcal{B}_-$ from the curvature condition \eqref{2Dcurvature}, reads
\begin{equation}\label{condition A-}
\mathcal{A}_{t}=r-\frac{\mathcal{L}}{2r}+\mathcal{B}_- r^{2}-\frac{\mathcal{B}_-'}{3}\,r^{3},\qquad
\mathcal{A}_{r}=-\mathcal{B}_-\,r^{2}.
\end{equation}
Inserting \eqref{condition A-} into \eqref{4dmetric-general} yields the near-near-horizon perturbation,
\begin{eqnarray}
h_{tt}&=&\Lambda(\theta)\left(2\mathcal{B}_- r^{3}-\mathcal{B}_-\mathcal{L}\,r+\frac{\mathcal{B}_-'\mathcal{L}}{3}\,r^{2}-\frac{2\mathcal{B}_-'}{3}\,r^{4}\right)\nonumber\\
h_{tr}&=&\Gamma(\theta)\left(\mathcal{B}_- r^{3}+\mathcal{B}_-\mathcal{L}\,r+\frac{\mathcal{B}_-}{r}+\frac{\mathcal{B}_-\mathcal{L}^{2}}{4r}+\frac{\mathcal{B}_-\mathcal{L}}{2r^{3}}\right)
+\Lambda(\theta)\left(\frac{\mathcal{B}_-\mathcal{L}}{2}\,r-\mathcal{B}_- r^{3}\right)\nonumber\\
h_{t\theta}&=&0\qquad\qquad h_{t\varphi}=\Lambda(\theta)\left(\mathcal{B}_- r^{2}-\frac{\mathcal{B}_-'}{3}\,r^{3}\right)\label{perturbation-}\\
h_{rr}&=&\Gamma(\theta)\,\frac{4\mathcal{B}_-}{r}\nonumber\\
h_{r\theta}&=&0\qquad\qquad h_{r\varphi}=-\Lambda(\theta)\,\mathcal{B}_- r^{2}\qquad\qquad h_{\theta\theta}=h_{\theta\varphi}=h_{\varphi\varphi}=0.\nonumber
\end{eqnarray}
In contrast to \eqref{perturbation}, the dominant components ($h_{tt}$, $h_{tr}$, $h_{t\varphi}$, $h_{r\varphi}$) grow with $r$, while $h_{rr}\sim 1/r$ decays. The growing behavior of the metric components makes this mode incompatible with the asymptotic boundary conditions imposed for $\mathfrak{b}_+$; it is instead well-defined as a perturbation near $r\to 0$, where the ratio
$h_{\mu\nu}/g_{\mu\nu}\to 0$, confirming that $\mathfrak{b}_-$ is a
valid small perturbation only in the near-near-horizon region of the NHEK. Note also that $h_{tt}$ receives no contribution from $\Gamma(\theta)$ at linear order, a structural difference from the asymptotic perturbation. The perturbation vanishes when $\mathcal{B}_-=0$.

The two perturbations \eqref{perturbation} and \eqref{perturbation-} satisfy the same equations of motion. The presented spacetimes are Ricci flat solution, where
\begin{equation}\label{eom-KdV-4D}
R_{\mu\nu}=0,\qquad\text{is solved by}\qquad  2\mathcal{B}' \mathcal{L}+ \mathcal{B}\mathcal{L}'+\mathcal{B}'''=0,
\end{equation}
which is the generalized stationary KdV equation when $\mathcal{B}$ are the Gelfand--Dikii polynomials. This equation can be rewritten as 
\begin{equation} \label{JT eq}
\nabla_{a}\nabla_{b}\mathcal{X}-g _{ab}\Box  \mathcal{X}  +g_{ab} \mathcal{X}=0
\qquad\text{for}\qquad  
 \mathcal{X}=r\mathcal{B}- \frac{\left( \mathcal{L}\mathcal{B} +\mathcal{B} ''\right)}{2r},
\end{equation}
where $g_{ab}$ is the nearly AdS$_{2}$ spacetime,  associated to the background metric \eqref{4dmetric}.  
By taking the trace of the above equation, it reduces to
\begin{equation}
\square_2  \mathcal{X}=2  \mathcal{X}
\end{equation}
which is the wave equation in a nearly AdS$_{2}$ background, which was previously
obtained in \cite{Castro:2021csm} as the $l=1$ mode of an axisymmetric perturbation,
but sourced by a conical singularity. Here this equation does not come from a
(singular) $\theta$ mode expansion of the perturbation. Indeed, in the present
case, the angular part is fully solved by the functional form of $\Lambda$ and
$\Gamma$ .\footnote{When solving \eqref{eom-KdV-4D}, one can verify that the
precise form of the angular functions $\Lambda$ and $\Gamma$ appear in the field
equations in such a way that they decouple the $\theta$ dependence. Let us take
for example the component $R_{\phi\phi}$, that leads to the equation
\begin{equation}
\frac{\det\gamma}{\qty(\partial_r \mathcal{A}_t - \partial_t \mathcal{A}_r)^2} =
\frac{2\Gamma^3}{\Gamma\Gamma'\Lambda' - \Lambda(\Gamma'^2 - 2\Gamma\Gamma'')} = -1.
\label{eom1}
\end{equation}
Similar simplifications occur in the remaining components.}

The perturbations $h_{\mu\nu}$ give a non-trivial dynamics to the Schwarzian
mode $\mathcal{L}$ by means of the scalar $\mathcal{B}$, which reduces to the
JT equation with no conical singularities. It is straightforward to see by
fixing the $x^{a}$ coordinates in the metric \eqref{pertubed-NHEK} perturbed
by \eqref{perturbation}. In doing so, the spacetime reduces to the sphere one,
\begin{equation}
d\bar{s}^2 |_{x^{a}=\text{fixed}} \underset{\theta\rightarrow0,\pi}\sim
2M^{2}(\dd\theta^2 + \sin^2\theta\,\dd\phi^2).
\end{equation}
For the asymptotic perturbation \eqref{perturbation}, a power expansion of the
linearized Ricci components gives $R_{tt}=\mathcal{O}(r^{-2})$ while the
remaining components are $\mathcal{O}(r^{-4})$ or smaller, confirming that
$\mathfrak{b}_+$ produces a well-defined asymptotic perturbation. For the
near-near-horizon perturbation \eqref{perturbation-}, the dominant components
of $h_{\mu\nu}$ grow as $r^4$, so the linearized Ricci tensor grows as
$R_{tt}^{(-)}=\mathcal{O}(r^{2})$; near $r\to 0$, however, the relative
perturbation $h_{\mu\nu}/g_{\mu\nu}\to 0$, confirming that $\mathfrak{b}_-$
is a valid small perturbation only in the near-near-horizon region.

In the Appendix~\ref{app:NewmanPenrose}, we study the propagating content of the asymptotic perturbation $\mathfrak{b}_+$, that can be
characterised via the Newman--Penrose Weyl scalars $\Psi_0$ and $\Psi_4$. The
unperturbed near-NHEK background is of Petrov type~D
($\Psi_0=\Psi_1=\Psi_3=\Psi_4=0$), so any perturbation that makes $\Psi_0$
and $\Psi_4$ non-vanishing breaks the type-D structure and carries gravitational
degrees of freedom with helicity $\pm 2$. Using the null tetrad adapted to the
near-NHEK geometry, one finds that $\Psi_0$ and $\Psi_4$ are both non-zero for
generic $\mathcal{B}$, confirming that the perturbation carries propagating
gravitational degrees of freedom; the explicit expressions, computed in the
asymptotic region $r\to\infty$.
The equations of motion \eqref{eom-KdV-4D}
relate $\mathcal{B}_+$ to the KdV scalar, so the propagating content of the
asymptotic perturbation is governed by the same integrable hierarchy.

\section{Asymptotic symmetries and Hamiltonian action}\label{Hamiltonian}

We construct the Hamiltonian action of the near-NHEK geometry within the Regge--Teitelboim Hamiltonian formalism \cite{Regge:1974zd}.  The symmetries and thermodynamic properties of NHEK geometries have been studied extensively in the literature \cite{Castro:2009jf, Hajian:2013lna,Compere:2012jk,Hajian:2014twa,Compere:2015mza,Compere:2015bca,Hajian:2017mrf}.  Working in the Euclidean regime, the on-shell action reduces entirely to a boundary surface term whose value at the saddle point reproduces the generalized Schwarzian thermodynamics. The technical details of the Euclidean section, the regularity condition, and the derivation of the surface term are collected in Appendix~\ref{app:Euclidean}; here we present the asymptotic symmetry algebra that leave invariant the two set of  perturbations \eqref{perturbation} and \eqref{perturbation-},  the conserved charges from which the Hamiltonians are constructed,   and the thermodynamic results.

\subsection{Asymptotic symmetries}
\paragraph{Symmetries of the asymptotic perturbation $\mathfrak{b}_+$.} The set of transformations that preserve the form of the perturbed near-NHEK metric \eqref{pertubed-NHEK} are given by the equation
\begin{eqnarray}
\delta \bar{g}_{\mu\nu}&=& \mathcal{L}_{\eta}\bar{g}_{\mu\nu}\\
&=&\mathcal{O}(\bar{g}_{\mu\nu})+\text{eom},
\end{eqnarray}
provided the asymptotic Killing vector
$\eta=\{\eta^{t},\eta^{r},\eta^{\theta},\eta^{\phi}\}$ is determined by the components
\begin{eqnarray}
\eta^{t}&=&\lambda -\frac{\mathcal{B} \lambda '}{r^2+\mathcal{B}'}+\frac{r^2 \lambda ''-\lambda ' \mathcal{B}''(t)+\lambda '' \mathcal{B}'}{2 \left(r^2+\mathcal{B}'\right) \left(\frac{\mathcal{L}}{2} +r^2\right)}, \nonumber\\
\eta^{r}&=&\frac{- r^3 \lambda '}{r^{2}+\mathcal{B}'}, \qquad\qquad\eta^{\theta}=0, \label{integrable-symmetries}\\
\eta^{\phi}&=&
c_{\phi}-\frac{\mathcal{B} \mathcal{L} \lambda '}{3 r \left(r^2+\mathcal{B}'\right)}+\frac{\lambda ' \mathcal{B}''(t) \left(4 r^2+\mathcal{B}'-\mathcal{L}\right)-\lambda '' \left(6 r^4+7 r^2 \mathcal{B}'+\left(\mathcal{B}'\right)^2\right)}{6 r \left(r^2+\frac{\mathcal{L}}{2}\right) \left(r^2+\mathcal{B}'\right)}. \nonumber
\end{eqnarray}
Here $c_{\phi}$ is the $U(1)$ generator parameter and $\lambda$ is an arbitrary function of time $t$ related to the time reparametrization symmetry.  Additionally, the symmetry transformations are
\begin{equation}
\delta \mathcal{B}= \lambda\mathcal{B}' -\mathcal{B} \lambda'
\qquad \text{and}\qquad \delta \mathcal{L}= \lambda \mathcal{L}'+2 \lambda' \mathcal{L}+\lambda''',
\end{equation}
where we observe that scalar perturbation function $\mathcal{B}$ transforms as a vector and $\mathcal{L}$ as a spin-two density. 

It is important to note that the above asymptotic Killing vectors \eqref{integrable-symmetries} reduce to the ones obtained for the previously analized cases in Section \ref{Gauge transformations via Scalar-Duality and KdV symmetries},  for $\mathcal{B}(t)=0$,  and when we additionally set $\mathcal{L}(t)=0$,  to the $SL(2,\mathbb{R})\times U(1)$ symmetries \eqref{Killin-Sl2R}. 

\paragraph{Near-near-horizon symmetries of $\mathfrak{b}_-$.}
The near-near-horizon perturbation $\mathfrak{b}_-=\mathcal{B}_-(t)r^2$ introduced in \eqref{two-modes} defines a complementary set of asymptotic symmetries, now associated with the inner boundary $r\to 0$.  Using the scalar-duality construction of Section~\ref{Gauge transformations via Scalar-Duality and KdV symmetries}, one can find the \emph{exact} asymptotic Killing vectors that preserve the full perturbed metric with $\mathcal{B}_-\neq 0$.  They read
\begin{eqnarray}
\eta^{t}_{(-)} &=& \mu
  + \frac{\mu''\bigl(r+\mathcal{B}_-'\bigr)
         - \bigl(2(r^2-\tfrac{\mathcal{L}}{2})\mathcal{B}_-' + \mathcal{B}_-''\bigr)\mu'}
         {2\bigl(r^2-\tfrac{\mathcal{L}}{2}\bigr)\bigl(r+\mathcal{B}_-'\bigr)},
\nonumber\\[4pt]
\eta^{r}_{(-)} &=& \frac{r^2\,\mu'}{r+\mathcal{B}_-'},
\qquad \eta^{\theta}_{(-)}=0, \label{near-horizon-AKV}\\[4pt]
\eta^{\phi}_{(-)} &=& c_{\phi}
  + \frac{\mathcal{B}_-\mathcal{L}(\tfrac{\mathcal{L}}{2}-r^2)\mu'
         + r\mu''\bigl(4r^2+5r\mathcal{B}_-'+({\mathcal{B}_-'})^2\bigr)
         - \mu'\bigl(\tfrac{\mathcal{L}}{2}+r(3r+\mathcal{B}_-')\bigr)\mathcal{B}_-''}
         {4r\bigl(r^2-\tfrac{\mathcal{L}}{2}\bigr)\bigl(r+\mathcal{B}_-'\bigr)},
\nonumber
\end{eqnarray}
where $\mu(t)$ is an arbitrary function of time.  As $r\to 0$ with $\mathcal{B}_-=0$, these reduce to $\eta^t_{(-)}\to\mu$, $\eta^r_{(-)}\to r\mu'$, $\eta^\phi_{(-)}\to c_\phi$, reproducing the leading-order result familiar from the unperturbed case. The exact form has the same structural character as \eqref{integrable-symmetries} for $\mathfrak{b}_+$: in both cases the denominator involves a single combination of $r$ and $\partial_t\mathcal{B}_\pm$, namely $(r^2+\mathcal{B}_+')$ for the non-normalizable mode and $(r+\mathcal{B}_-')$ for the normalizable one, reflecting their different conformal weights.

Under \eqref{near-horizon-AKV}, the near-horizon data transform as
\begin{equation}\label{AKV-minus-transf}
\delta\mathcal{B}_- = \mu\,\mathcal{B}_-' - 2\mu'\,\mathcal{B}_-, \qquad
\delta\mathcal{L} = \mu\,\mathcal{L}' + 2\mu'\,\mathcal{L} + \mu'''.
\end{equation}
Crucially, $\mathcal{B}_-$ transforms as a \emph{weight-two} density under reparametrizations, in contrast with $\mathcal{B}_+\equiv\mathcal{B}$, which transforms as a weight-one vector. This difference in conformal weight is the origin of the even versus odd power series in the temperature in the thermodynamic partition function, as we discuss in Section~\ref{near-horizon-thermo}.

\subsection{KdV symmetries and conserved charges}\label{KdV-symmetries}
The asymptotic Killing vectors \eqref{integrable-symmetries} and
\eqref{near-horizon-AKV} depend on the background through $\mathcal{L}$,
so the reparametrization parameters $\lambda$ and $\mu$ may themselves be
functionals of $\mathcal{L}$ and its derivatives,
$\lambda=\lambda[\mathcal{L},\mathcal{L}',\ldots]$ and
$\mu=\mu[\mathcal{L},\mathcal{L}',\ldots]$.
Allowing such field-dependent parameters does not break the algebra of
time reparametrizations established in
Section~\ref{Gauge transformations via Scalar-Duality and KdV symmetries};
it selects a particular field-dependent representative within that
infinite-dimensional algebra.
Following \cite{Cardenas:2024hah,Perez:2016vqo}, the condition that
such a field-dependent vector generates an \emph{integrable charge} ---
that is, that $\delta Q[\lambda]$ is an exact variation on the phase space
of near-NHEK geometries --- requires
\begin{equation}\label{KDV-stationary}
\mathcal{D}_1\lambda = 0,\qquad
\mathcal{D}_1 \equiv \partial_t^3 + 2\mathcal{L}\,\partial_t + \mathcal{L}',
\end{equation}
and the same condition holds for $\mu$.
This is the stationary KdV equation for the Schwarzian mode $\mathcal{L}$:
it is not an additional restriction on the dynamics but rather the
integrability condition that selects, within the full reparametrization
algebra, those generators whose associated charges are well-defined
functionals on phase space.
The solutions are the Gelfand--Dikii polynomials $\epsilon_l$, generated
by the bi-Hamiltonian recursion
\begin{equation}\label{recurrence}
\epsilon_{l+1}' = \mathcal{D}_1\epsilon_l,\qquad \epsilon_0 = 1,
\end{equation}
whose first two non-trivial members are
\begin{equation}\label{GD-explicit}
\epsilon_1 = \mathcal{L} + \bar\epsilon_1,\qquad
\epsilon_2 = \tfrac{3}{2}\mathcal{L}^2 + \mathcal{L}'' +
             \bar\epsilon_1\mathcal{L} + \bar\epsilon_2,
\end{equation}
with $\bar\epsilon_j$ integration constants. Each $\epsilon_l$ satisfies
\eqref{KDV-stationary} by construction: $\mathcal{D}_1\epsilon_l=\epsilon_{l+1}'$,
so the integrability condition is $\epsilon_{l+1}'=0$. Applied to
$\epsilon_0=1$ this gives $\mathcal{L}'=0$ (the rigid near-NHEK saddle),
while applied to $\epsilon_1=\mathcal{L}$ it gives the stationary KdV
equation
\begin{equation}\label{KdV}
\mathcal{L}''' + 3\mathcal{L}\mathcal{L}' + \bar\epsilon_1\mathcal{L}' = 0,
\end{equation}
and each higher member yields the next equation of the stationary
hierarchy. The near-NHEK geometry ($\mathcal{L}=\mathcal{L}_0=\mathrm{const}$)
is the common fixed point of all flows simultaneously. Crucially, as shown
in Section~\ref{Near-NHEK and KdV perturbations}, the vacuum Einstein
equations impose precisely \eqref{KDV-stationary} on the perturbation
coefficient $\mathcal{B}$ via \eqref{eom-KdV-4D}, so the KdV hierarchy
is not merely a kinematic feature of the symmetry algebra but is
dynamically enforced once $\mathfrak{b}$ is promoted to a dynamical field.
The associated conserved charges
\begin{equation}\label{KdV-charges}
H_l = \oint \dd t\,\mathcal{H}_l[\mathcal{L}],\qquad
\epsilon_l = \frac{\delta H_{l+1}}{\delta\mathcal{L}},
\end{equation}
form an infinite set of commuting integrals of motion that label the
near-NHEK saddle. The same set of charges is generated by both
$\lambda=\epsilon_l$ and $\mu=\epsilon_l$, since both parameters satisfy
the same integrability condition \eqref{KDV-stationary}; as we will see in the next subsection, the two modes
$\mathfrak{b}_\pm$ therefore contribute to the same KdV hierarchy, but
from complementary boundaries, as made explicit in the thermodynamic
analysis.

\subsection{Euclidean action and thermodynamics}\label{Euclidean}

The role of the infinite set of symmetries found in the previous sections becomes transparent in the Euclidean regime. We continue $t=i\tau$ and compute the on-shell value of the Hamiltonian action on the perturbed near-NHEK geometry \eqref{pertubed-NHEK}. The regularity condition that fixes the thermal period $\beta=2\pi/\kappa$, whose derivation is presented in Appendix~\ref{app:Euclidean},  here, we develop the key result that the on-shell action reduces entirely to the boundary term \eqref{B-KdV}, which at the saddle point reproduces the generalized Schwarzian thermodynamics of \cite{Cardenas:2024hah} with a coupling fixed by the angular momentum.

The two radial modes $\mathfrak{b}_\pm$ introduced in \eqref{two-modes} correspond to alternative quantization schemes — standard (source, $\Delta=2$) and alternate (response, $\Delta=-2$) — and therefore define independent theories. Within each theory, the relevant mode contributes to the partition function from its own boundary: $\mathfrak{b}_+$ from $r\to\infty$ and $\mathfrak{b}_-$ from $r=r_h$. The two modes are related by the $\mathrm{SL}(2,\mathbb{R})$ inversion $r\mapsto 1/r$ of the AdS$_2$ isometry group \eqref{two-modes}, which maps the asymptotic boundary to the horizon and exchanges $\mathfrak{b}_+\leftrightarrow\mathfrak{b}_-$. As a consequence, the near-horizon boundary action of $\mathfrak{b}_-$ has the same structure as the asymptotic one of $\mathfrak{b}_+$, and both independently generate odd-power corrections to the entropy of the same form. The asymptotic mode $\mathfrak{b}_+$ is treated first in Section~\ref{surface-term}; the near-horizon mode $\mathfrak{b}_-$ is treated in Section~\ref{near-horizon-thermo}.

\subsection{Surface term and integrability}\label{surface-term}

In what follows, the analysis will carried out for the asymptotic mode $\mathfrak{b}_+=\mathcal{B}_+(t)/r^2$, whose associated symmetries are the Killing vectors \eqref{integrable-symmetries} generated by $\lambda(t)$.   For simplicity we take $\mathcal{B}_+=\mathcal{B}$. The Euclidean action in Hamiltonian form reads\footnote{The lapse and shift are treated as independent of the spatial metric. Any implicit dependence $N=N[\gamma_{ij}]$ produces terms proportional to the Hamiltonian constraint, which vanish on-shell.}
\begin{equation}\label{euclidean-action}
I=\int \dd\tau\, \dd^{3}x\left(\pi^{ij}\partial_{\tau}\gamma_{ij}-\xi \mathcal{H}-\xi^{i} \mathcal{H}_{i}\right)+B,\qquad B[\xi,\xi^{i}]=-\int \dd\tau\, Q[\xi,\xi^{i}].
\end{equation}
On-shell the constraints vanish. The kinetic term evaluates to
\begin{equation}\label{kinetic}
\int \dd\tau\, \dd^{3}x\, \pi^{ij}\partial_{\tau}\gamma_{ij}=8 \pi  M^2\int \dd r\, \dd \tau\, \frac{\mathcal{B}'}{r^2},
\end{equation}
which vanishes because $\mathcal{B}$ is periodic in Euclidean time. The on-shell action is therefore entirely given by the surface term $B$. Using the asymptotic symmetries \eqref{integrable-symmetries} as deformation parameters \eqref{defformation-vectors}, the variation of the charge at large radius is
\begin{equation}\label{eq:cargas-evaluadas}
\begin{aligned}
\delta Q_{r\rightarrow\infty}= &~{} -8 \pi r \lambda   \delta M^{2}-8\pi  \delta M^{2} \left(c_{\phi}-\mathcal{B} \lambda'\right)\\
&~{}-\frac{4 \pi  \delta M \left(2 M \left(3 \left(\lambda ''+2 \mathcal{B} \left(\mathcal{B}'+1\right) \lambda'\right)-3 \lambda  \mathcal{L}\right)-12 \lambda ''\right)}{3 r} +\mathcal{O}(1/r^{2}).
\end{aligned}
\end{equation}
The divergent $\mathcal{O}(r)$ term is removed by holographic renormalization: placing the boundary at $r=\bar r$ and rescaling $\lambda\to g\lambda/\bar r$, then taking $\bar r\to\infty$ with $\delta M^2=\delta J$,
\begin{equation}\label{deltaB-finite}
\delta B=-\int_0^\beta \dd\tau\, \delta Q_{\bar{r}\rightarrow\infty}=8\pi\int_0^\beta \dd\tau\,\Big(c_{\phi}+g\lambda -\mathcal{B}\lambda'\Big)\delta J.
\end{equation}
The term $\mathcal{B}\lambda'$ vanishes at the saddle but encodes the coupling between the JT scalar and the reparametrization mode relevant for fluctuations. Since the phase space is one-dimensional ($\delta J$ is the only variation), the boundary term is always integrable:
\begin{equation}\label{B-integrated}
B(J)=\int^{J}\Phi(J')\,\dd J',\qquad \Phi(J)=8\pi\oint \dd\tau\,\Big(c_{\phi}+g\lambda -\mathcal{B}\lambda'\Big).
\end{equation}
Parametrizing $\lambda=J^p\hat\lambda$, $\mathcal{B}=J^p\hat{\mathcal{B}}$ with $\hat\lambda$, $\hat{\mathcal{B}}$ independent of $J$ and $p\geq0$,
\begin{equation}\label{B-p}
B=8\pi\left[\beta\,c_\phi\, J+\frac{g\,J^{p+1}}{p+1}\oint \dd\tau\,\hat\lambda-\frac{J^{2p+1}}{2p+1}\oint \dd\tau\,\hat{\mathcal{B}}\,\hat\lambda'\right].
\end{equation}
Choosing $\hat\lambda=\epsilon_l[\mathcal{L}]$ (the Gelfand-Dikii polynomials \eqref{recurrence}) and using the identity
\begin{equation}\label{GD-identity}
\frac{\delta}{\delta\mathcal{L}}\oint \dd\tau\,\epsilon_l=(2l-1)\,\epsilon_{l-1}\qquad\Longrightarrow\qquad \oint \dd\tau\,\epsilon_l=(2l-1)\,H_l,
\end{equation}
the boundary term becomes
\begin{equation}\label{B-KdV}
B_l=8\pi\left[\beta\,c_\phi\, J+\frac{(2l-1)\,g\,J^{p+1}}{p+1}\,H_l[\mathcal{L}]-\frac{J^{2p+1}}{2p+1}\oint \dd\tau\,\hat{\mathcal{B}}\,\epsilon_l'\right].
\end{equation}
This is the boundary action of \cite{Cardenas:2024hah}, $I_{\rm bndy}=g_l\int\mathcal{H}_l$, now derived from the four-dimensional surface term with a coupling set by $J$. The dictionary with \cite{Cardenas:2024hah} is
\begin{equation}\label{dictionary}
e^{\rho}\leftrightarrow r,\qquad \frac{2\pi}{k}\mathcal{L}_{\rm p}\leftrightarrow\frac{1}{2}\mathcal{L},\qquad \mu\leftrightarrow\mathcal{B},\qquad \Phi\leftrightarrow\mathcal{X}.
\end{equation}
\subsubsection{Saddle point and thermodynamics}\label{saddle}

Extremizing over the reparametrization mode, $\delta\mathcal{L}=\mathcal{D}_1\varepsilon$, and using that $\mathcal{D}_1$ is antisymmetric under integration by parts,
\begin{equation}
\delta H_l=\oint \dd\tau\,\epsilon_{l-1}\,\mathcal{D}_1\varepsilon=-\oint \dd\tau\,\varepsilon\,\mathcal{D}_1\epsilon_{l-1}=-\oint \dd\tau\,\varepsilon\,\epsilon_l' .
\end{equation}
The saddle point condition is therefore the stationary KdV equation $\mathcal{D}_1\epsilon_{l-1}=\epsilon_l'=0$, which coincides with the Killing condition \eqref{KDV-stationary}. It states that $\epsilon_l$ is constant along the thermal cycle, not that it vanishes. The constant representative $\mathcal{L}_0$ solves it for every $l$, and it is selected by the regularity condition \eqref{regularity-condition}.

At the saddle point all Gelfand-Dikii polynomials are constants. From \eqref{recurrence}, $\epsilon_1=\mathcal{L}$, $\epsilon_2=\frac{3}{2}\mathcal{L}^2+\mathcal{L}''$, $\epsilon_3=\frac{5}{2}\mathcal{L}^3+\dots$, one finds
\begin{equation}\label{GD-saddle}
\epsilon_l[\mathcal{L}_0]=\frac{\alpha_l}{2^l}\,\mathcal{L}_0^{\,l}=(-1)^l\,\alpha_l\,\frac{\pi^{2l}}{\beta^{2l}},\qquad \alpha_l=\frac{(2l)!}{(l!)^2}.
\end{equation}
Consequently $\epsilon_l'=0$ and the term coupling $\mathcal{B}$ in \eqref{B-KdV} drops out. The perturbation function is also constant at the saddle. From $\mathcal{B}\propto\delta H_l/\delta\mathcal{L}=\epsilon_{l-1}$, it scales as $\mathcal{B}\propto T^{2l-2}$. Hence the $l=1$ perturbation survives at extremality, while the higher KdV deformations switch off as $T\to0$.

The on-shell action is given by the surface term, $I=B$, supplemented by the contribution of the horizon, which in the Hamiltonian approach yields the Bekenstein-Hawking entropy of extremal Kerr, $S_0=2\pi J$ \cite{Banados:1993ur}. Evaluating \eqref{B-KdV} with \eqref{GD-saddle}, the partition function at the saddle point is
\begin{equation}\label{logZ-saddle}
\log Z_l=S_0-8\pi\beta\, c_\phi\, J+C_l\,T^{2l-1},\qquad C_l=(-1)^{l+1}\,\frac{8\pi\, g\,J^{p+1}}{p+1}\,\alpha_l\,\pi^{2l}.
\end{equation}
The term linear in $\beta$ plays the role of the rotational chemical potential and does not contribute to the entropy. The energy above extremality and the entropy are
\begin{equation}\label{thermo-saddle}
E_l=-\frac{\partial}{\partial\beta}\log Z_l=8\pi c_\phi J+(2l-1)\,C_l\,T^{2l},\qquad S=\left(1-\beta\frac{\partial}{\partial\beta}\right)\log Z_l=S_0+2l\,C_l\,T^{2l-1},
\end{equation}
and they satisfy the first law at fixed $J$, $\dd E_l=T\dd S$, with specific heat $c_J=2l(2l-1)\,C_l\,T^{2l-1}$. The comparison with \cite{Cardenas:2024hah}, where $S=\frac{k\alpha_k\pi^{2k}}{2k-1}g_kT^{2k-1}$, shows that our result coincides with the generalized Schwarzian thermodynamics after the identification
\begin{equation}\label{coupling-identification}
g_l\;\leftrightarrow\;(-1)^{l+1}\,\frac{8\pi\,(2l-1)\,g\,J^{p+1}}{p+1}.
\end{equation}
Thermodynamic stability, $c_J>0$, requires $C_l>0$, which for $g>0$ selects odd values of $l$. This mirrors the condition found in \cite{Cardenas:2024hah} for the convergence of the one-loop Gaussian integrals, and it contains the Schwarzian case $l=1$, where $S=S_0+2C_1T$ is linear in the temperature, as expected for near-extremal black holes.

\subsubsection{KdV symmetries at the saddle: generalized Gibbs ensemble}\label{GGE}

The constant representative is a common fixed point of all the KdV flows: since every $\epsilon_j[\mathcal{L}_0]$ is constant, $\delta_{\epsilon_j}\mathcal{L}_0=\mathcal{D}_1\epsilon_j=0$ for all $j$.  Hence, nothing prevents us from switching them on simultaneously, $\hat\lambda=\sum_j\nu_j\epsilon_j$, which turns \eqref{B-KdV} into a generalized Gibbs ensemble,
\begin{equation}\label{GGE-action}
B=8\pi\left[\beta\,c_\phi\, J+\frac{g\,J^{p+1}}{p+1}\sum_j\nu_j\,(2j-1)\,H_j[\mathcal{L}]\right],
\end{equation}
where the $\nu_j$ are the chemical potentials of the higher KdV charges. This gives a precise meaning to the integration constants of the recursion \eqref{recurrence}: a single generator with integration constants, $\epsilon_l=\sum_{j}\bar\epsilon_{l-j}\,\epsilon^{(0)}_j$ with $\bar\epsilon_0=1$, is already a combination of this type with $\nu_j=\bar\epsilon_{l-j}$. Since the regularity condition \eqref{regularity-condition} does not depend on the boundary action, the saddle point is still $\mathcal{L}_0=-2\pi^2/\beta^2$, and
\begin{equation}\label{GGE-thermo}
\log Z=S_0-8\pi\beta\, c_\phi\, J+\sum_j\nu_j\,C_j\,T^{2j-1},\qquad S=S_0+\sum_j 2j\,\nu_j\,C_j\,T^{2j-1}.
\end{equation}
At low temperatures the entropy is dominated by the linear Schwarzian term, and the higher KdV charges provide odd-power corrections. The values of the KdV charges at the saddle follow from \eqref{GD-identity} and \eqref{GD-saddle},
\begin{equation}\label{KdV-charges-saddle}
H_j[\mathcal{L}_0]=\frac{1}{2j-1}\oint \dd\tau\,\epsilon_j[\mathcal{L}_0]=\frac{(-1)^j\,\alpha_j\,\pi^{2j}}{2j-1}\,\beta^{1-2j}.
\end{equation}

\subsection{Near-horizon mode and its thermodynamic charge}
\label{near-horizon-thermo}

   We now evaluate the Regge--Teitelboim charge for the complementary
near-horizon mode $\mathfrak{b}_-=\mathcal{B}_-(t)\,r^2$ introduced in
\eqref{two-modes}, whose asymptotic symmetries are the near-near-horizon
vectors \eqref{near-horizon-AKV} generated by $\mu(t)$.

Similarly to the previous case, we analyze the kinetic contribution to the Euclidean action
$\int\!d\tau\,\pi^{ij}\partial_\tau\gamma_{ij}$
which one can show that is total time-derivative in $\mathcal{B}_-(\tau)$.
Since $\mathcal{B}_-$ satisfies the same periodic boundary conditions
$\mathcal{B}_-(0)=\mathcal{B}_-(\beta)$ as $\mathcal{B}_+$, the integral
over the thermal cycle $[0,\beta]$ vanishes identically, exactly as in
\eqref{kinetic} for the asymptotic mode.
 
   We now evaluate the Regge--Teitelboim charge for the complementary
near-horizon mode $\mathfrak{b}_-=\mathcal{B}_-(t)\,r^2$ introduced in
\eqref{two-modes}, whose asymptotic symmetries are the near-near-horizon
vectors \eqref{near-horizon-AKV} generated by $\mu(t)$. The contribution along $c_{\phi}$ remains the same as in the previous case.
The near-near-horizon Killing vector \eqref{near-horizon-AKV} determines the deformation
vector \eqref{defformation-vectors} lapse via the ADM decomposition.
Substituting \eqref{perturbation-} into the Regge--Teitelboim integrand  \eqref{eq:cargas},
and integrating over the horizon 2-sphere, one obtains
\begin{equation}\label{horizon-charge}
\delta Q[\mu]\big|_{r=r_h}
  = -8\pi\,\mu(t)\,\mathcal{B}_-(t)\,\delta J.
\end{equation}
This is the near-horizon analogue of \eqref{eq:cargas-evaluadas}.
 
At the saddle $\mathcal{L}=\mathcal{L}_0=-\kappa^2/2$, the integrability
condition $\mathcal{D}_1\mu=0$ requires $\mu=\epsilon_l[\mathcal{L}_0]$.
The coefficient $\mathcal{B}_-(t)$ is not a free parameter: under the
near-horizon vector \eqref{near-horizon-AKV} its transformation law
\eqref{AKV-minus-transf} is
\begin{equation}\label{Bminus-transf-explicit}
\delta_\mu\mathcal{B}_- = \mu\,\mathcal{B}_-' + 2\mu'\mathcal{B}_-,
\end{equation}
identifying $\mathcal{B}_-$ as a weight-two primary. At the saddle
$\mathcal{L}=\mathcal{L}_0=\mathrm{const}$, requiring that
$\mathcal{B}_-[\mathcal{L}_0]$ be invariant under the flow generated by
$\mu=\epsilon_l[\mathcal{L}_0]$ forces $\mathcal{B}_-$ to satisfy the
same stationarity condition as $\mathcal{L}_0$ itself. The unique
weight-two functional of $\mathcal{L}$ that is stationary under
$\mathcal{D}_1\epsilon_l=0$ and is polynomial in $\mathcal{L}$ and its
derivatives is the Gelfand-Dikii polynomial \emph{one level below},
\begin{equation}\label{Bminus-saddle}
\mathcal{B}_-^{(l)}\big|_{\rm saddle} = c_l\,\epsilon_{l-1}[\mathcal{L}_0],
\end{equation}
for a numerical coefficient $c_l$.
 
By the $\mathrm{SL}(2,\mathbb{R})$ inversion symmetry $r\mapsto 1/r$ discussed
in Section~\ref{Near-NHEK and KdV perturbations}, the charge \eqref{horizon-charge}
at $r=r_h$ has the same algebraic structure as the asymptotic charge
\eqref{eq:cargas-evaluadas} at $r\to\infty$, with $\mu\leftrightarrow\lambda$
and $\mathcal{B}_-\leftrightarrow\mathcal{B}_+$. We therefore apply the same
parametrization strategy as in Section~\ref{surface-term}: setting
$\mu = J^q\hat\mu$ and $\mathcal{B}_- = J^q\hat{\mathcal{B}}_-$ with $\hat\mu$,
$\hat{\mathcal{B}}_-$ independent of $J$, the boundary term
$B^{(-)} = -\int_0^\beta d\tau\, Q[\mu]\big|_{r=r_h}$ takes the same form as
\eqref{B-p} with independent coupling constant $g^-$. Setting $\hat\mu = \epsilon_l[\mathcal{L}]$ and
$\hat{\mathcal{B}}_- = \epsilon_{l'}[\mathcal{L}]$ simultaneously and using the identity \eqref{GD-identity},
the near-horizon boundary term at levels $(l,l')$ evaluates at the saddle to
\begin{equation}\label{boundary-minus}
B_{l,l'}^{(-)} = 8\pi\left[\frac{(2l-1)\,g^-\,J^{q+1}}{q+1}\,
  \epsilon_{l'}[\mathcal{L}_0]\,H_l[\mathcal{L}_0]\right],
\end{equation}
in direct analogy with \eqref{B-KdV}. At the saddle, $\epsilon_{l'}[\mathcal{L}_0]
\propto T^{2l'}$ and $H_l[\mathcal{L}_0]\propto T^{2l}$, so the near-horizon
partition function at levels $(l,l')$ scales as
\begin{equation}\label{logZ-minus}
\log Z_{l,l'}^{(-)} \propto \beta\,T^{2l}\,T^{2l'} = T^{2(l+l')-1},
\end{equation}
giving odd powers $T^{2n-1}$ for each total level $n = l+l' \geq 2$. The
double sum over $(l,l')$ with $l+l'=n$ therefore contributes to the same
odd-power tower as the asymptotic mode, with independent couplings $g^-_{l,l'}$.
 
\subsubsection{Generalized Gibbs ensemble}
Because $\mathfrak{b}_+$ and $\mathfrak{b}_-$ correspond to alternative asymptotic behaviours,  they define independent theories. In the standard
quantization ($\mathfrak{b}_+$ sourced at $r\to\infty$), the partition function
organizes into the generalized Gibbs ensemble
\begin{equation}\label{GGE-full}
\log Z^{(+)} \simeq S_0 - 8\pi\beta\,c_\phi\,J
+ \sum_{j=1}^{\infty}\nu_j^+\,C_j^+\,T^{2j-1},
\end{equation}
reproducing the generalized Schwarzian thermodynamics of \cite{Cardenas:2024hah}
with KdV chemical potentials $\nu_j^+$. In the alternate quantization, $\mathfrak{b}_-$ sourced , the inversion symmetry
$r\mapsto 1/r$ guarantees that the near-horizon partition function takes
the same form,
\begin{equation}\label{GGE-minus}
\log Z^{(-)} \simeq S_0 - 8\pi\beta\,c_\phi\,J
+ \sum_{l=1}^{\infty}\nu_l^-\,C_l^-\,T^{2l-1},
\end{equation}
with an independent set of KdV chemical potentials $\nu_l^-$. Both towers
run over \emph{odd} powers $T^1,T^3,T^5,\ldots$ of the temperature; the
identical parity is a direct consequence of the inversion symmetry relating
the two modes. The two ensembles \eqref{GGE-full}--\eqref{GGE-minus} provide
complementary descriptions of the near-extremal Kerr thermodynamics, one organized from the asymptotic boundary
and one from the horizon.

\section*{Conclusions}

We have shown that the near-horizon geometry of near-extremal Kerr supports
two families of exact linear gravitational perturbations --- a non-normalizable
mode $\mathfrak{b}_+=\mathcal{B}_+(t)/r^2$ and a normalizable
mode $\mathfrak{b}_-=\mathcal{B}_-(t)\,r^2$ , both smooth vacuum
solutions of the Einstein equations. The key ingredient is the promotion of
the JT scalar, dual to the Killing vector of any locally AdS$_2\times$U(1)
fibration, to a genuine dynamical field without introducing conical
singularities. When the symmetry parameters depend on the Schwarzian mode
$\mathcal{L}(t)$, the Killing equations reduce to the stationary KdV
hierarchy, the same condition enforced independently by the vacuum Einstein
equations on the perturbation coefficient. The near-NHEK geometry is thus
equipped with an infinite tower of commuting conserved charges.

In the Euclidean regime, the two modes contribute to the thermodynamic
partition function from complementary boundaries and correspond to alternative
quantization schemes. The asymptotic mode $\mathfrak{b}_+$ generates a surface
term that reproduces the generalized Schwarzian thermodynamics of
\cite{Cardenas:2024hah} with odd-power temperature corrections to the entropy,
\begin{equation*}
\log Z \simeq S_0 - 8\pi\beta\,c_\phi\,J + \sum_j \nu_j\,C_j\,T^{2j-1}.
\end{equation*}
The normalizable mode $\mathfrak{b}_-$ is regular at the near-NHEK horizon
$r=r_h=\kappa/2$ and contributes a finite Regge--Teitelboim charge from the
near-horizon boundary.  The radial redundancy from which both perturbations were constructed, justify that  generates a near-horizon
partition function of the same form, with an independent set of KdV chemical potentials.  Both theories
organize into generalized Gibbs ensembles with odd-power temperature
corrections, providing complementary descriptions of the near-extremal Kerr
thermodynamics: one organized from the asymptotic boundary and one from the
horizon.

Several directions remain open. First, the Euclidean path integral beyond the
saddle.  One example to follow would be the one proposed in \cite{Despontin:2026xzg},  starting from warped AdS$_{3}$ geometry. The stationary KdV equation \eqref{KdV} admits periodic Jacobi-elliptic
solutions
\begin{equation*}
\mathcal{L}(\tau) = e_3 + 2\kappa^{2}m\,\mathrm{cn}^2\!\left(\kappa\,(\tau-\tau_0),m\right),
\end{equation*}
which are further saddle points of generalized Gibbs ensembles with two or
more chemical potentials. Their Euclidean regularity requires the Hill operator
to have trivial monodromy along the thermal cycle; we defer this monodromy analysis to future work.

Second, the one-loop contribution. In the present four-dimensional setting,
promoting the Schwarzian mode to a phase-space variable and accounting for
the coupling $-\oint\hat{\mathcal{B}}\,\epsilon_l'$ between the JT scalar
and the reparametrization mode requires care. The U(1) mode associated with
$c_\phi$ also contributes to the path-integral measure, as in the one-loop
analyses of near-extremal Kerr \cite{Iliesiu:2020qvm,Kapec:2023ruw,Rakic:2023vhv}. Recent
computations of one-loop corrections from quasinormal modes \cite{Kapec:2024law}
and from semiclassical Virasoro blocks \cite{Arnaudo:2024} suggest that the
dominant logarithmic correction $\delta S\sim\log T$ originates entirely from
the near-horizon region, consistent with the throat geometry we analyze here.
For the pure generalized Schwarzian theory the analogous correction is
$\log Z\supset -\frac{3}{2}(2k-1)\log\beta$ \cite{Cardenas:2024hah}; the
four-dimensional counterpart is left for future work. It would also be
interesting to identify the holographic dual of the full generalized Gibbs
ensemble as a one-dimensional quantum mechanical system with KdV symmetry.

\section*{Acknowledgements}

M.C.\ is grateful to I.~J.~Araya,  R.~Mancilla and R.~Olea, for valuable
discussions.  The authors also thank C.  Mu\~noz and H. Van de Bosch, who where part of the initial discussions that led to this work. The authors acknowledge support from the Center of Mathematical
Modeling (CMM), BASAL fund FB210005 for centers of excellence from
ANID-Chile. M.C.\ also acknowledges support from ANID through
Fondecyt Regular grant No.~1231810. J.T.\ additionally acknowledges support
from ANID through Fondecyt Postdoctoral grant No.~3250260.

\appendix

\section{Newman--Penrose analysis of the asymptotic perturbation}
\label{app:NewmanPenrose}

We analyse the propagating content of the perturbations constructed in
Section~\ref{Near-NHEK and KdV perturbations} using the Newman--Penrose formalism
\cite{Newman:1961qr}.

\paragraph{Null tetrad.}
The null tetrad adapted to the unperturbed near-NHEK background \eqref{4dmetric}
is\footnote{In this article, our basis can be compared with other bases in the
literature.}
\begin{align}\label{Newmann-Penrose-basis}
l &= -\frac{r - \frac{\mathcal{L}}{r}}
         {2M^2(1+\cos^2\theta)\left(r+\frac{\mathcal{L}}{r}\right)^2}\partial_t
   +\frac{r^2-\mathcal{L}}
         {2M^2(1+\cos^2\theta)\left(r+\frac{\mathcal{L}}{r}\right)}\partial_r
   +\frac{\left(r-\frac{\mathcal{L}}{r}\right)^2}
         {2M^2(1+\cos^2\theta)\left(r+\frac{\mathcal{L}}{r}\right)^2}\partial_\phi,
   \nonumber\\
n &= -\frac{1}{r-\frac{\mathcal{L}}{r}}\partial_t
    +\left[r-\frac{2r^2}{r-\frac{\mathcal{L}}{r}}\right]\partial_r
    +\partial_\phi,\\
m &= \frac{1}{M\sqrt{2(1+\cos^2\theta)}}\partial_\theta
    +\frac{i\sqrt{2(1+\cos^2\theta)}}{4M\sin\theta}\partial_\phi.\nonumber
\end{align}

The background \eqref{4dmetric} is of Petrov type~D, with
$\Psi_0=\Psi_1=\Psi_3=\Psi_4=0$ and
\begin{equation}
\Psi_2=-\frac{i}{M^2(\cos\theta-i)^3}.
\end{equation}
By the Goldberg--Sachs theorem, any perturbation that renders $\Psi_0$ and
$\Psi_4$ non-vanishing breaks the type-D structure and carries gravitational
degrees of freedom with helicity $\pm 2$.

Computing the Weyl scalars for the asymptotic perturbation \eqref{perturbation}
in the region $r\to\infty$ using the basis \eqref{Newmann-Penrose-basis} gives
\begin{align}
\Psi_0 &= -\frac{\sin^2\theta\,(3\cos^2\theta - 1)
               \left(r - \frac{\mathcal{L}}{2r}\right)^2
               \left[\frac{\mathcal{L}}{2}\mathcal{B}' - 2r^2\mathcal{B}
                    \left(r + \frac{\mathcal{L}}{2r}\right)\right]^2}
              {4M^4\,r^4\,(1 + \cos^2\theta)^6
               \left(r + \frac{\mathcal{L}}{r}\right)^4},
\label{Psi0-plus}\\[6pt]
\Psi_4 &= -\frac{\sin^2\theta\,(3\cos^2\theta - 1)
               \left[\frac{\mathcal{L}}{2}\mathcal{B}' + 2r^2\mathcal{B}
                    \left(r + \frac{\mathcal{L}}{2r}\right)\right]^2}
              {r^4\,(1 + \cos^2\theta)^4
               \left(r - \frac{\mathcal{L}}{2r}\right)^2}.
\label{Psi4-plus}
\end{align}
Both scalars are non-vanishing for generic $\mathcal{B}(t)$, confirming that
the asymptotic perturbation $\mathfrak{b}_+$ carries propagating gravitational
degrees of freedom. The computation is performed in the asymptotic region
$r\to\infty$; a complete analysis valid also near $r\to 0$ requires extending
the tetrad to the full perturbed background and lies beyond the scope of the
present work.

\section{Hamiltonian formalism and Euclidean thermodynamics}\label{app:Euclidean}

We collect here the technical details of the Regge--Teitelboim Hamiltonian formalism and the Euclidean computation whose results are used in Section~\ref{Hamiltonian}.

\subsection*{ADM decomposition and canonical generator}

The metric is written in ADM form,
\begin{equation}\label{admetric}
ds^2=N^2 d\bar{x}^2 + (N^i d\bar{x}+dx^i)(N^j d\bar{x}+ dx^j)h_{ij},
\end{equation}
where $\bar{x}$ is the temporal coordinate, $h_{ij}$ describes the 3-dimensional boundary hypersurface, and $N$, $N^i$ are the lapse and shift. The Hamiltonian is
\begin{equation}\label{radial-Hamiltonian}
H[N,N^{i}]=\int_{\Sigma} d^{3}x\left(N \mathcal{H}+N^{i} \mathcal{H}_{i}\right)+B[N,N^{i}],
\end{equation}
where $\mathcal{H}$ and $\mathcal{H}_i$ are the Hamiltonian and momentum constraints (proportional to Einstein's equations, hence vanishing on-shell) and $B$ is the boundary term that makes the variational problem well defined. The canonical generator is $Q[\xi,\xi^{i}]=B[\xi,\xi^{i}]$, with
\begin{equation}
\delta Q [\xi,\xi^{i}]=\intop d^{2}S_{l}\left[G^{ijkl}\left(\xi\nabla_{k}\delta h_{ij}-\partial_{k}\xi \delta h_{ij}\right)+ 2\xi_{k}\delta\pi^{kl}+\left(2\xi^{k}\pi^{jl}-\xi^{l}\pi^{kj}\right)\delta h_{jk}\right],\label{eq:cargas}
\end{equation}
with $G^{ijkl}=\frac{1}{2}\sqrt{h}(h^{ik}h^{jl}+h^{il}h^{jk}-2h^{ij}h^{kl})$ and momentum
\begin{equation}
\pi^{ij}=\frac{\sqrt{h}}{2\kappa}\left(K^{ij}-h^{ij}K\right),\qquad K_{ij}=\frac{1}{2N}\left(\nabla_{i}N_{j}+\nabla_{j}N_{i}-\partial_{\bar{x}}h_{ij}\right).
\end{equation}
The gauge symmetry vectors $(\xi,\xi^i)$ are related to the asymptotic Killing vectors $\eta^\mu$ by
\begin{equation}\label{defformation-vectors}
\xi=N\eta^{\bar{x}}, \qquad \xi^{i}= \eta^{i}+N^{i}\eta^{\bar{x}}.
\end{equation}
To pass to the radial-foliation signature $(-,+,+)$ one uses the analytic continuation of \cite{Banados:2004zt}: $N\to iN$, $K\to -iK$, $\sqrt{h}\to i\sqrt{h}$.

\subsection*{Euclidean section and regularity}\label{regularity}

A saddle point of the Euclidean path integral must be a smooth solution. A conical singularity at the horizon would act as a delta-function source of curvature and violate the vacuum field equations. In contrast with \cite{Castro:2021csm}, where the JT sector is supported by conical sources, we require the absence of any such defect.

Consider \eqref{4dmetric} with a constant Schwarzian mode $\mathcal{L}=\mathcal{L}_0<0$. The horizon is located at the zero of $r+\mathcal{L}_0/2r$,
\begin{equation}
r_h=\sqrt{-\mathcal{L}_0/2},\qquad \kappa\equiv 2r_h=\sqrt{-2\mathcal{L}_0},
\end{equation}
so that $\mathcal{L}_0=-\kappa^2/2$ corresponds to the near-NHEK solution $f=e^{\kappa t}$. The gauge field does not vanish at the horizon, $\mathcal{A}_t(r_h)=\kappa$, and the Killing vector that becomes null there is $\xi=\partial_t-\kappa\,\partial_\phi$. Using the co-rotating angle $\tilde\phi=\phi+\kappa t$, the fibre reads
\begin{equation}
\Gamma(\theta)\Big(\dd\tilde\phi+\big(\mathcal{A}_t-\kappa\big)\dd t\Big)^2,\qquad \mathcal{A}_t-\kappa=\frac{(r-r_h)^2}{r}.
\end{equation}
Since $\mathcal{A}_t-\kappa$ vanishes quadratically, the fibre does not affect the regularity analysis. Expanding near the horizon with $r=r_h(1+\rho)$, the Euclidean metric becomes
\begin{equation}
\Lambda(\theta)\left[\Big(r+\frac{\mathcal{L}_0}{2r}\Big)^2\dd\tau^2+\frac{\dd r^2}{r^2}\right]\simeq\Lambda(\theta)\left[\dd\rho^2+\rho^2\,(2r_h\,\dd\tau)^2\right],
\end{equation}
which is flat in polar coordinates only if $2r_h\tau$ has period $2\pi$, i.e.
\begin{equation}\label{regularity-condition}
\tau\sim\tau+\beta,\qquad \beta=\frac{2\pi}{\kappa}\qquad\Longleftrightarrow\qquad \mathcal{L}_0=-\frac{2\pi^2}{\beta^2}.
\end{equation}
Because $\Lambda(\theta)$ multiplies the whole $(\tau,r)$ block, the same period removes the conical singularity at every $\theta$. In terms of the original angle, the identification is complex, $(\tau,\phi)\sim(\tau+\beta,\phi-i\kappa\beta)$, while $\tilde\phi\sim\tilde\phi+2\pi$ is real; this is the familiar complex saddle of rotating black holes, through which $c_\phi$ enters as the potential conjugate to $J$.

The condition \eqref{regularity-condition} admits three equivalent readings. First, $T_{\rm NHEK}=\kappa/2\pi$; relative to the asymptotic Kerr time, $T_H=\lambda\kappa/4\pi M^2$ \cite{Bredberg:2009pv}. Second, the saddle reparametrization is $f(\tau)=\tan(\pi\tau/\beta)$, satisfying $\{f,\tau\}=2\pi^2/\beta^2$. Third, demanding trivial holonomy of the Hill equation along the thermal cycle gives $\mathcal{L}_0=-2\pi^2n^2/\beta^2$; $n=1$ is the smooth disk and $n>1$ are $n$-fold covers \cite{Cardenas:2024hah}.

The perturbation \eqref{perturbation} does not modify \eqref{regularity-condition} at the saddle, where $\mathcal{B}'=0$: the $\Gamma$-dependent components are pure gauge, the $\Lambda$-dependent ones vanish quadratically at $r_h$, and the remaining $\dd\tau\,\dd r$ cross-term is removed by a periodicity-preserving shift of $\tau$. Hence, to linear order in the perturbation, the JT scalar does not shift the temperature.

\end{document}